\documentclass[a4paper]{jpconf}
\usepackage{graphicx}
\usepackage{mathabx}

\newcommand{\EA}[1]{\mbox{10$^{#1}$}}

\newcommand{\et}{\mbox{$\eta$}}
\newcommand{\etp}{\mbox{$\eta'$}}

\newcommand{\elp}{\mbox{$e^+$}}
\newcommand{\elm}{\mbox{$e^-$}}

\begin{document}
\title{Tests of the fundamental symmetries in  $\eta$ meson decays}

\author{Andrzej Kupsc$^1$ and Andreas Wirzba$^2$}

\address{$^1$ Department of Physics and Astronomy,
    Uppsala University, Box 516, 75120 Uppsala, Sweden and
High Energy Physics Department,
    The Andrzej Soltan Institute for Nuclear Studies,
    Hoza 69, 00-681, Warsaw, Poland}
\address{$^2$ Institute for Advanced Simulation and 
   J\"ulich Center for Hadron Physics,
   Institut f\"ur Kernphysik,
    Forschungszentrum J\"ulich,
    D-52425 J\"ulich, Germany
    }
\ead{$^1$ andrzej.kupsc@physics.uu.se, $^2$ a.wirzba@fz-juelich.de}

\begin{abstract}
Patterns of chiral symmetry violation and tests of the conservation of
the fundamental $C,\ P$ and  $CP$ symmetries are key physics issues in
studies of the  $\pi^0$, $\eta$ and $\eta'$ meson  decays. These tests
include  searches  for  rare  or  forbidden decays  and  searches  for
asymmetries among the decay  products in the not-so-rare decays.  Some
examples for the rare decays are $\eta\to2\pi$, $\eta\to 4\pi^0$ ($CP$
tests),   decays  into  an   odd  number   of  photons   ({\it  e.g.},
$\eta\to3\gamma$) and the  decay $\eta\to\pi^0e^+e^-$ ($C$ tests).  In
the conversion  decay $\eta\to\pi^+\pi^-e^+e^-$, the  asymmetry in the
distribution of  the angle between  the two-pion and  two-lepton decay
planes  allows to search  for $CP$  violation in  a flavor--conserving
process  beyond the  CKM mechanism  which  is not  constrained by  the
limits on the  neutron dipole moment.  In addition,  the Dalitz decays
and  the   decays  into  a lepton--antilepton  pair  are   sensitive  to
contributions from a vector  boson responsible for the annihilation of
hypothetical  light dark matter particles.   The muon  $g-2$  and the
branching ratio  for the  $\pi^0\to e^+ e^-$  decay are  currently the
observables  where  hints  of  a  deviation from  the  Standard  Model
predictions are  reported.  The  experimental studies of  the $\pi^0$,
$\eta$  and $\eta'$  meson decays  are  carried out  at four  European
accelerator  research  facilities:  KLOE/KLOE-2 at  DAFNE  (Frascati),
Crystal Ball at MAMI (Mainz),  WASA at COSY (J\"ulich), Crystal Barrel
at ELSA (Bonn).
\end{abstract}

%%%%%%%%%%%%%%%%%%%%%%%%%%%%%%%%%%%%%%%%%%%%%%%%
\section{Introduction}

The $\pi^0$,  $\eta$ and $\eta'$  mesons belong to the  longest living
hadrons decaying predominantly via electromagnetic or strong processes
that  are eigenstates  of parity  ($P$) and  charge  conjugation ($C$)
operators.  They therefore allow for studies of the (non)-conservation
of  the $C$,  $P$ and  $CP$ symmetries  in strong  and electromagnetic
interactions.  In  the case  of the $\pi^0$  and $\eta$ meson  all the
strong  and electromagnetic decays  are either  forbidden or  at least
severely   suppressed.    Nevertheless,    there   exist   many   open
(energetically allowed) final states  for the $\eta$ and $\eta'$ meson
decays, such that a variety of symmetry tests is possible.

The    most    probable   decays    of    the    $\eta$   meson    are
\cite{Nakamura:2010zzi}: the  second order electromagnetic  decay into
two  photons  (BR $\approx$  39\%),  which  is  driven by  the  chiral
(triangle) anomaly, and the  isospin-violating decays into three pions
($\eta\to\pi^0\pi^0\pi^0$ @BR $\approx$ 33\%, $\eta\to\pi^+\pi^-\pi^0$
@BR $\approx$  23\%), which are  mainly due to the  difference between
the $u$  and $d$ quark masses since  electromagnetic contributions are
suppressed.  The first order radiative decay $\eta\to\pi^+\pi^-\gamma$
(BR $\approx$ 5\%) is also induced by the chiral anomaly.

Any radiative  decay with one  or more photons  in the final  state is
accompanied by a  process where the photon converts  internally into a
lepton-antilepton  pair~\cite{Dalitz:1951aj}.   The conversion  decays
allow to study  the {\it  transition form factors} describing the structure
of the interaction  region. The form factors are  also measured in the
$\gamma^*\gamma^*\to\eta$   processes  in  $e^+e^-$   colliders.   The
precise knowledge  of the transition  form factors of the  $\pi^0$ and
$\eta$ mesons  is needed  for the calculations  of the  Standard Model (SM)
contributions to  the muon  $g-2$ and to  the rare $\pi^0$  and $\eta$
meson decays into a lepton--antilepton pair.

%%%%%%%%%%%%%%%%%%%%%%%%%%%%%%%%%%%%%%%%%%%%%%%%

\section{Tests of CP symmetry}

Most of the tests of the $CP$ symmetry in $\eta$ decays are modeled as
flavor conserving counterparts of the corresponding $K_L$ decay modes.
The  straightforward test  is to  search  for $P$  and $CP$  violating
$\eta$ decays into  two pions.  The best experimental  limits are from
KLOE\,\cite{Ambrosino:2004ww},   BR$({\eta\to\pi^+\pi^-})<  1.3  \cdot
10^{-5}$, and GAMS-4$\pi$\,\cite{Blik:2007ne}, BR$(\eta\to\pi^0\pi^0)<
3.5  \cdot  10^{-4}$.   The  theoretical  branching  ratios  are  very
small\,\cite{Jarlskog:2002zz}: in  the SM, the
decays are $G_F^2$ processes additionally suppressed by a cancellation
of  nearly  equal  terms  leading  to  the  branching  ratio  estimate
BR$(\eta\to  \pi\pi)  \leq 2  \cdot  10^{-27}$.   The branching  ratio
generated  by a $CP$  violation in  the extended  Higgs sector  of the
electroweak    theory   is   listed    in\,\cite{Jarlskog:1995gz}   as
$\mathrm{BR}(\eta             \to             \pi^+\pi^-)            =
2\mathrm{BR}(\eta\to\pi^0\pi^0)\leq 1.2  \cdot 10^{-15}$.  If strong
interaction physics  via the  $\theta$ term were  the culprit  for the
$CP$  violation, the  empirical bound\,\cite{Nakamura:2010zzi}  on the
electric  dipole  moment of  the  neutron  would  limit the  pertinent
branching ratio to BR$(\eta\to  \pi\pi)_\theta \leq 3 \cdot 10^{-17}$.
In fact, as argued  by Gorchtein\,\cite{Gorchtein:2008pe} the bound on
the electric  dipole moment  of the  neutron can be  used to  impose a
limit  on  the branching  ratio  $\mathrm{BR}(\eta\to\pi\pi) \leq  3.5
\cdot 10^{-14}$ no matter what the underlying $CP$ violating mechanism
would turn out to be.

The search  for the barely  (energetically) allowed $\eta$  decay into
4$\pi^0$,   proposed  by  Nefkens\,\cite{Nefkens:1994hc},   see also  Ref.\,\cite{Prakhov:2000xm},
represents  a new  kind  of $CP$  test  with no  analog  in the  $K_L$
system. If all the final pions are in relative $s$ waves, the decay is
forbidden by $P$  and $CP$ invariance.  Due to the very low excess energy
of 7.9\,MeV this assumption seems  to be plausible.  However, $CP$ can
be conserved if the final state involves higher partial waves than $s$
or $p$  waves.  Therefore it  is interesting to consider  the analogue
$\eta'\to 4 \pi^0$  decay where the phase space is  big enough for the
four pions to group into two $d$-wave pion pairs which can couple to a
combined angular  momentum of $0^+, 1^+,\cdots,  4^+$.  If furthermore
the two pairs are then in a relative $p$-wave state, {\it i.e.} if the
orbital  angular  momentum  between  the  pairs  is  $1^-$,  the  full
four-pion  system   can  be  in  a  pseudoscalar   $J^{PC}  =  0^{-+}$
state.

\begin{figure}[htb]
\begin{center}
 \includegraphics[width=13cm]{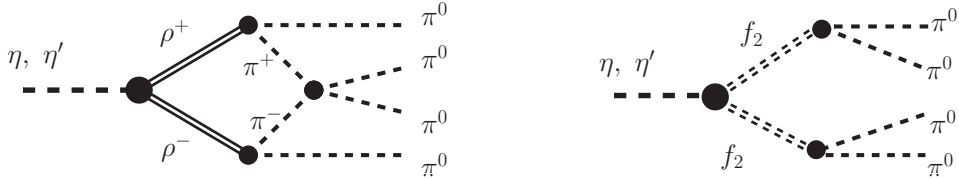}
\end{center}
\caption{\label{fig:4pion}(Left) An  example of an  anomaly-driven 
  pion-loop/rescattering diagram contributing  to the  $CP$ allowed
  $\eta/\eta'\to  4\pi^0$  decay.  Note   that  to  the  left  of  the
  rescattering  vertex  the $\pi^\pm\pi^0$  pairs  resulting from  the
  vector  mesons are  $p$-wave states,  whereas  to the  right of  the
  vertex the  upper and  lower (or inner and outer) $\pi^0\pi^0$ pairs  have to  be $d$-wave
  states.  Nevertheless,  the   overall  angular  momentum  is  always
  $J^{PC}=0^{-+}$  because of  the relative  $p$-wave  orbital angular
  momentum between  the two  vector mesons and  between the  upper and
  lower (inner and outer) $\pi^0\pi^0$ pairs. \   
  (Right) Direct decay $\eta^{(\prime)}\! \to\!
  f_2  f_2\!\to\!   (\pi^0\pi^0)  (\pi^0\pi^0)$  where   the  two  virtual
  isoscalar $d$-wave  bosons $f_2$ are emitted in  a relative $p$-wave
  state. Both processes are of order ${\cal O}(p^{10})$\,\cite{GuoKubisWirzba2011}.}
\end{figure}

Since  an odd number  of (quasi)-Goldstone bosons  is involved,
the decay is driven by the chiral anomaly or, more precisely, by a higher order  term (than the
fourth order of the Wess-Zumino-Witten action) of the  intrinsic-parity-odd sector of 
Chiral Perturbation Theory, see {\it e.g.}  \cite{Bijnens:2001bb,Ebertshauser:2001nj}. 
Note that the simple flavor structure  together with the Bose symmetry of the pions excludes
a direct contribution of the pentangle anomaly. For the same reason, also the decay
path $\eta^{(\prime)}\to \rho^0\rho^0\to (\pi^0  \pi^0) (\pi^0\pi^0)$ is forbidden.
In fact,  as shown in \cite{GuoKubisWirzba2011}, even 
direct contributions of  chiral order ${\cal O}(p^6)$ Lagrangians tabulated in
\cite{Bijnens:2001bb,Ebertshauser:2001nj} and, moreover, direct and indirect ${\cal O}(p^{8})$ contributions 
can be excluded,
such that chiral order ${\cal O}(p^{10})$ processes come into play:
{\it e.g.},  a triangle-anomaly  decay into two rho mesons (which, in fact, is already of ${\cal O}(p^6)$) 
plus (at  least) one additional $d$-wave pion loop/rescattering of order ${\cal O}(p^4)$ or  a
direct  process via two isoscalar $f_2$ tensor-mesons,  $\eta^{(\prime)}  \to  f_2  f_2 \to  (\pi^0\pi^0)
(\pi^0 \pi^0)$, which is of chiral order ${\cal O}(p^{10})$\,\cite{GuoKubisWirzba2011}, see Figure~\ref{fig:4pion}.

%%%%%%%%%%%%%%%%%%%%%%%%%%%%%%%%%%%%%%%%%%%%%%%%%%%%%%%
%%%%%%%%%%%%%%%%%%%%%%%%%%%%%%%%%%%%%%%%%%%%%%%%%%%%% 

Using naive phase  space arguments the $CP$ allowed  4-pion decays can
be estimated -- at least close to  threshold -- to scale as $ (p\, r_0
)^{1+3+3 +  2\cdot 2+2\cdot 2 +2\cdot  1}= (p r_0)^{17}$  where $p$ is
the averaged  three-momentum modulus of a $\pi^0$  in the $\eta,\eta'$
center-of-mass frame and $r_0 \sim 1/m_\rho$ is the effective range of
the interaction between two  pions expressed by the $\rho$-meson mass.
This  predicts a  tiny ratio  of the  partial  widths ${\Gamma(\eta\to
  4\pi^0)}/{\Gamma(\eta'\to   4   \pi^0)}   \le  10^{-16}$.   The
estimates of the branching ratio  limits, however, have to involve the
comparison  with  other  decay  channels ({\it  e.g.}  $\eta\to  \pi^0
\pi^0\pi^0$, $\eta\to \pi^+\pi^-  l^+ l^-$, $\eta'\to \pi^0\pi^0 \eta$
etc.)  and  are more uncertain: $\le 10^{-10}$  for the  $\eta$ and
(with more caveat because of  the increased distance to the threshold)
$\le 10^{-8}$ for the $\eta'$. A measurement of a non-zero BR for
these $\eta/\eta' \to 4\pi^0$  decays above the specified limits would
probably be a signal for a $CP$ violating process.

Experimentally  the  main advantage  of  the  decay  is the  very  low
background  from  the direct  pion  production  if  the experiment  is
carried out  in $\eta$ meson production reactions  close to threshold.
In  fact,  the  experimental   limit  obtained  by  the  Crystal  Ball
collaboration\,\cite{Prakhov:2000xm},   namely  the  $BR   <  6.9\cdot
10^{-7}$,  is the most  sensitive result  on any  of the  $\eta$ meson
decays.  As  mentioned above, an interesting possibility  is to search
for the $\eta'\to 4\pi^0$ decay.   The experimental limit on the BR is
$5\cdot 10^{-4}$ from the GAMS-2000 spectrometer\,\cite{Alde:1987jt}.
 
The    search    for     linearly    polarized    photons    in    the
$\eta\to\pi^+\pi^-\gamma$   decay    was   proposed   by    Geng   and
collaborators\,\cite{Geng:2002ua}.      As     this     process     is
flavor-conserving and  involves an extra  photo-production vertex, the
sensitivity of this  $CP$ test is not constrained  by the experimental
limits on the $\eta\to \pi\pi$ decays or the electric dipole moment of
the neutron or by $CP$ violating processes in the flavor changing kaon
and $B$-meson sectors.   The practical realization of this  idea is to
investigate  the $\eta\to \pi^+\pi^- e^+  e^-$ process\,\cite{Gao:2002gq}.
The violation of  the CP symmetry in the  $\eta\to \pi^+\pi^- e^+ e^-$
decay would manifest itself as an angular asymmetry between the pionic
and the  lepton-antilepton decay planes.   In the case of  the $K_L\to
\pi^+\pi^-     e^+     e^-$      decay     such     an     asymmetry\,
\cite{Cabibbo:1965zz,Pais:1968zz,Bijnens:1994me} is driven by standard
CKM    mechanism    and    it     was    observed    by    the    KTeV
collaboration\,\cite{AlaviHarati:1999ff}       and       the      NA48
collaboration\,\cite{Lai:2003ad}.  In the case  of the $\eta$ system a
possible  mechanism  leading  to   such  an  asymmetry  could  be  the
interference between the usual  $CP$ allowed magnetic $M_1$ transition
(driven by the chiral  anomaly) and a $CP$ violating flavor-conserving
electric dipole operator\,\cite{Geng:2002ua} which is sensitive to the
strange quark  content of the decaying  $\eta$ meson and  which is not
constrained by SM physics.   The experimental bound on this asymmetry,
measured by the KLOE collaboration, is still compatible with zero, $A_\phi =
(-0.6\pm 3.1)\cdot 10^{-2}$~\cite{Ambrosino:2008cp}.

%%%%%%%%%%%%%%%%%%%%%%%%%%%%%%%%%%%%%%%%%%%%%%%%

\section{Tests of C symmetry}
\label{sec:TestC}

There   exists   only   limited   knowledge  of   the   $C$-symmetry
(non-)conservation  in strong  and  electromagnetic interactions.  Any
$\eta$ decay into  {\em neutrals} with an {\em  odd} number of photons
in the  final state  will not conserve  $C$.  The simplest  example of
this class is the decay  into three photons which, however, is heavily
suppressed:  each photon pair  has to  involve at  least two  units of
angular    momentum   because   of    gauge   invariance    and   Bose
symmetry\,\cite{Dicus:1976my,Grosse:2001ei}; the case of a photon pair
with  total  angular  momentum   zero  is  excluded,  since  it  would
correspond to a radiative $0\to  0$ transition, which is forbidden for
a real  photon\,\cite{Sakurai:1964q}, while the case of  a photon pair
with  total   angular  momentum  $J=1$   is  in  conflict   with  Bose
symmetry\,\cite{Heiliger:1993ja}.   The present  upper limit  from the
KLOE experiment is  BR$(\eta\to\gamma\gamma\gamma) < 1.6\cdot 10^{-5}$
90\% CL~\cite{Aloisio:2004di}.

Also the $\eta$ decays into $\pi^0$ mesons and an odd number of direct
photons belong  to the above-mentioned class.   The simplest prototype
would  be  the  $\eta\to  \pi^0\gamma$  decay  which,  however,  as  a
radiative  $0\to  0$  transition  with  a real  photon  is  absolutely
forbidden by angular momentum conservation\,\cite{Sakurai:1964q}.  The
experimental bound  on this decay from the  Crystal Ball collaboration
is       $\mathrm{BR}(\eta\to      \pi^0\gamma)\leq       9      \cdot
10^{-5}$\,\cite{Nefkens:2005ka}.  The next simplest cases are then the
decays with  more than one  $\pi^0$ but only  one photon in  the final
state: $\eta\to\pi^0\pi^0\gamma$ and $\eta\to 3\pi^0\gamma$.  The best
branching ratio limits  for these decays come again  from the Crystal Ball
experiment   and   are   $5\cdot   10^{-4}$  and   $6\cdot   10^{-5}$,
respectively\,\cite{Nefkens:2005dp}.

\begin{figure}[tbh]

\vspace{3mm}
\begin{center}
 \includegraphics[width=14cm]{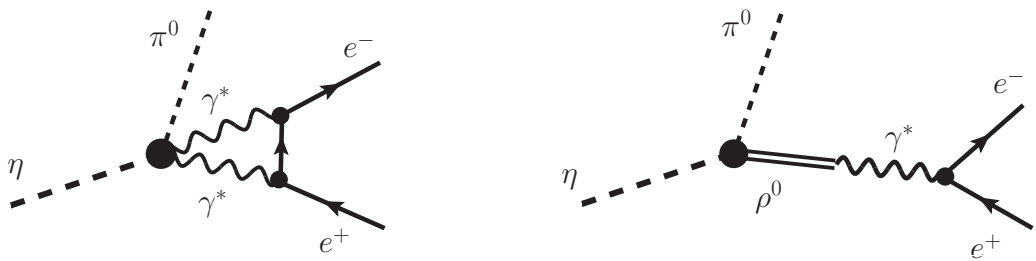}
\end{center}
\caption{\label{fig:pi0ee}  (Left)  Fourth-order  electromagnetic  
 $C$-conserving transition: $\eta\!\to\!\pi^0\gamma^*\gamma^*\to\pi^0 e^+
  e^-$.  (Right)  Diagram  for   a  possible  $C$-violating  process:
  $\eta\to\pi^0\rho\to\pi^0 e^+ e^-$.}
\end{figure}

In  addition,  $\eta$ decays  into  neutrals  plus  an odd  number  of
dilepton  pairs  in  the  final   state  as,  {\it  e.g.},  the  decay
$\eta\to\pi^0 e^+  e^-$ may  be used for  tests of  charge conjugation
invariance.   The  main SM  contribution  to  this  process (shown  in
Figure~\ref{fig:pi0ee})  comes  from  the $C$-conserving exchange  of  two
intermediate photons  ($\eta\to\pi^0\gamma^*\gamma^*\to\pi^0e^+ e^-$ )
with a branching ratio of about $10^{-8}$ \cite{Cheng:1967zza}.
The
process with  one virtual photon $\gamma^*$ in  the intermediate state
is forbidden  by $C$ symmetry and,  moreover, has to  have a vanishing
transition form factor in the on-shell limit, since -- as stated above
-- the  direct  $\eta\to\pi^0\gamma$ vertex  is  forbidden by  angular
momentum conservation\,\cite{Sakurai:1964q}.

At present,  the empirical upper  limit for the $BR(\eta\to\pi^  0 e^+
e^-)$ is equal to $4.5\cdot 10^{-5}$ 90\% CL, which comes from an 1975
experiment\,\cite{Jane:1975nt}.   The  data  used for  obtaining  this
limit  were  analyzed under  assumption  of  a  constant decay  matrix
element and  a cut  $M(e^+e^-)>140$ MeV was  applied.  For  the analog
decay $\eta\to  \pi^0 \mu^+ \mu^-$,  there exists a  similar branching
ratio      limit,      $BR(\eta\to\pi^0\mu^+\mu^-)<5\cdot     10^{-6}$
\cite{Dzhelyadin:1980ti}.    Furthermore,   also   the   single-photon
contributions  to  the  $\eta'$  decays  into  a  $\pi^0$  or\  $\eta$
pseudoscalar and a dilepton pair ($e^+ e^-$ or $\mu^+\mu^-$) belong to
the class of $C$-violating processes.

Finally,  the  $C$  invariance  can  be  tested  also  in, {\it  e.g.},
$\eta\to\pi^+\pi^-\gamma$ and  $\eta\to\pi^+\pi^-\pi^0$ decays where it
can manifest  itself as an  asymmetry in the energy  distributions for
$\pi^+$  and $\pi^-$  mesons in  the rest  frame of  the  $\eta$ meson
\cite{Lee:1965aax,Bernstein:1965hj}.     The   asymmetries    in   the
$\eta\to\pi^+\pi^-\gamma$  and  $\eta\to\pi^+\pi^-\pi^0$  decays  were
investigated  in  a few  experiments  in   the 1970s  \cite{Gormley:1970qz,
  Layter:1972aq,  Jane:1974es, Jane:1974mk,  Thaler:1972ax}.   For the
$\eta\to\pi^+\pi^-\pi^0$ asymmetries there exist recent limits from
the     KLOE      collaboration     with     sensitivity     $10^{-3}$
\cite{Ambrosino:2008ht}.

%%%%%%%%%%%%%%%%%%%%%%%%%%%%%%%%%%%%%%%%%%%%%%%%

\section{Dark force searches}

The rare decays  of a neutral pseudoscalar meson  $P$ ($\pi^0$, $\eta$
or  $\eta'$) into  one dilepton  pair ($e^+e^-$  or,  if kinematically
allowed,  $\mu^+\mu^-$) belong  to the  most interesting  processes of
low-energy hadron physics, since the SM calculation predicts minuscule
tree-level contributions (via a virtual $Z$ boson), such that a window
of opportunity for the admixture  of physics beyond the standard model
might open up as the rates are small.  In fact, the relative branching
ratio      BR$(P\to      l^+l^-)$/BR$(P\to     \gamma\gamma)      \sim
(\frac{\alpha}{\pi} \frac{m_l}{m_P})^2$ is  not only suppressed by two
powers  of the  fine-structure  constant $\alpha$  resulting from  the
coupling of {\em two} virtual photons to  the dilepton pair, but also by the
helicity  mismatch of  the two  outgoing leptons  -- indicated  by the
squared ratio of the lepton mass $m_l$ to the pseudoscalar mass $m_P$.
Recently the KTeV collaboration determined  a new precise value of the
branching  ratio BR$(\pi^0\to  e^+ e^-)=  (7.49\pm  0.29\pm 0.25)\cdot
10^{-8}$  which  exceeds the  up-to-date  theoretical calculations  by
Dorokhov                            {\it                            et
  al.}\,\cite{Dorokhov:2007bd,Dorokhov:2008cd,Dorokhov:2009xs}   by  3
standard deviations.

Kahn  and collaborators\,\cite{Kahn:2007ru}  suggested  as a  possible
explanation  of this excess  the tree-level  exchange of  an off-shell
neutral vector boson $U$ (pioneered by Fayet\,\cite{Fayet:1980rr,Fayet:1980ss})
of  mass $m_U$\ $\sim$\,(10 -- 100)\,MeV. The
latter is  of the  type proposed  in the light  dark matter  models by
Boehm                              {\it                             et
  al.}\,\cite{Boehm:2003hm,Boehm:2002yz,Boehm:2003bt,Boehm:2004gt}
 (see also Ref.\,\cite{Fayet:2004bw})
 to mediate the annihilation reaction $\chi \chi \to e^+ e^-$ of a neutral
scalar dark matter  particle $\chi$ of (1 --  10)\,MeV mass, such that
the  excess positrons  produced  in this  annihilation reaction  could
account for the bright 511\,keV  line emanating from the center of the
Galaxy\,\cite{Kahn:2007ru}.
 
In order to explain the mismatch between experiment and theory for the
branching    ratio    BR$(\pi^0\to     e^+e^-)$,    Kahn    {\it    et
  al.}\,\cite{Kahn:2007ru} assumed  that the neutral  vector meson $U$
would couple  to the $u$ and $d$  quark fields of the  $\pi^0$ and the
$e^+e^-$  dilepton  pair  via  the  axial-vector  components  $g_A^u$,
$g_A^d$ and $g_A^e$, respectively, where -- for simplicity -- a common
axial  coupling $g_A\equiv  g_A^u-g_A^d\equiv g_A^e$  of the  order of
$g_A  = (2.0  \pm 0.5)  \cdot 10^{-4}  \cdot m_U/(10\,{\rm  MeV})$ was
fitted.  In fact, the computation of the partial width of this process
is modeled  according to the  analog SM tree-level process  $\pi^0 \to
{Z}^\ast \to e^+  e^-$, where the $Z$-boson mass  $m_Z$ is replaced by
the much lighter mass $m_U$ and where the weak coupling is replaced by
$g_A^{u,d,e}$.  Assuming that the octet axial-vector quark-coupling is
of  the  same  order as  above,  the  $U$  boson contribution  to  the
branching ratio  BR$(\eta\to e^+e^-)$ is about $10^{-9}$,  which is of
the       same       order       as       the       estimates       of
Dorokhov\,\cite{Dorokhov:2007bd,Dorokhov:2008cd,Dorokhov:2009xs}    and
much smaller  than the experimental  bound $2.7 \cdot 10^{-5}$  of the
CELSIUS/WASA  collaboration\,\cite{Berlowski:2008zz}.   The same  fit,
however,  predicts a  contribution  of order  $2.0  \cdot 10^{-5}$  to
BR$(\eta\to \mu^+\mu^-)$ which is  nearly an order of magnitude larger
than  the  measured  value   BR$(\eta\to  \mu^+\mu^-)  \sim  (5.7  \pm
0.9)\cdot   10^{-6}$\,\cite{Abegg:1994wx},  unless   the  axial-vector
coupling of the  $U$ meson to the muon is smaller  than $g_A^e$ or the
octet  axial-vector quark  coupling is  smaller than  $g_A^u-g_A^d$ or
both. This  is of course a  limitation in the predictive  power of the
$U$ boson exchange mechanism.

Another variant of  a dark matter $U$ gauge  bosons,
the $a_\mu$,  
 was    suggested    by    Reece    and
Wang\,\cite{Reece:2009un}
(see also the  work by
Fayet\,\cite{Fayet:1986rh,Fayet:1989mq,Fayet:1990wx,Bouchiat:2004sp}):
here  the additional $U(1)_d$  `dark' gauge
boson couples to  the SM $U(1)$ gauge boson by  a gauge kinetic mixing
term ${\cal  L}_{\rm kin-mix}  = -2 \epsilon  F^{\mu\nu} F_{\mu\nu}^d$
with a  strength $\epsilon \sim  10^{-3}$ or less.  The  dark $U(1)_d$
group is assumed  to be spontaneously broken by  the introduction of a
dark higgs $h_d$  field which acquires a GeV  scale vacuum expectation
value,  such  that $m_U  \sim  1$\,MeV  --  few\,GeV.  Under  a  field
redefinition to  the standard massless  photon, the $U$  boson couples
{\em  vectorially}  to  all  SM  charged  fields  as  $\epsilon  a_\mu
J^{\mu}_{\rm  EM}$,  where  $J^\mu_{\rm   EM}$  is  the  pertinent  SM
electromagnetic current  of the particle.  In  addition, the $U$-boson
also couples  to SM weak neutral currents;  however, the corresponding
coupling is suppressed by a factor of order $m_U^2/m_Z^2$.  Therefore,
this $U$ boson  variant has very small axial  couplings to SM matter
fields and cannot serve as  a candidate for fitting the missing excess
in the $\pi^0\to e^+e^-$ decay.

First searches  for $U$ bosons of the  Reece and Wang type  are on the
way.  The relevant channels are, {\it e.g.}, the decays $\phi \to \eta
U,\  U \to  e^+ e^-$  and $\eta  \to  \gamma U,\  U \to  e^+ e^-$.   A
signature for  a $U$  boson would  be a peak  in the  pertinent Dalitz
decay $\phi \to  \eta e^+ e^-$ or $\eta \to  \gamma e^+e^-$, while the
background  for  such searches  is  the  standard  Dalitz decay.   The
searches for $U$ bosons in  the $\phi\to e^+e^- \eta$ conversion decay
are  carried  out by  KLOE  collaboration\footnote{S.  Giovannella  in
  DISCRETE2010  proceedings.}.

%With 
%700\,pb$^{-1}$ 
%data analyzed  in the $ \phi\to e^+e^- \eta,\ \eta\to\pi^+\pi^-\pi^0$ channel, an exclusion 
%bound for the above-mentioned $\epsilon$-parameter of the Reece-Wang-type $U$ boson could be reported:  $\epsilon < 3 \cdot 10^{-3}$ at 95\% CL in the invariant mass range 25\,MeV $< M_{ee} < $425\,MeV; note that systematics are not yet included.
%The alternative decay path $\eta\to \gamma\gamma$,  which might increase statistics, is still under investigation.
% In addition, a possible Higgs'-Bremsstrahlung process, suggested in \cite{Batell:2009yf}, 
% \[ e^+ e^- \to U' \to U h_d \to 
%\mu^+ \mu^-  + \mbox{`invisible'}\,,
%\] 
% which applies for the case $m_{h_d} < m_U$, is also under investigation.

Finally, as  the above mention  experiments will all aim  at searching
for a  narrow peak on a  large conversion background,  one should also
mention  an  attractive  alternative   for  $U$  boson  searches:  the
($C$)-forbidden  $\eta$  decays,  especially the  decay  $\eta\to\pi^0
e^+e^-$, where both the  conventional and the $C$-violating background
are suppressed for low $M(e^+e^-)$.  From the background point of view
it is an ideal process to search  for a low energy $U$ boson. From the
theory point  of view,  however, one has  to cope with  the additional
suppression by either the violation of $C$ symmetry, which applies to
the Reece-Wang  $U$ boson and  to the Boehm  {\it et al.}  case under
vector coupling, or by a violation  of parity $P$ in the Boehm {\it et
  al.} case under axial-vector coupling.

%%%%%%%%%%%%%%%%%%%%%%%%%%%%%%%%%%%%%%%%%%%%%%%%

\section{Experiments}

The experimental  programs for  $\eta$ and $\eta'$  decays are  on the
agenda both at  $e^+e^-$ colliders  and hadro-  or  photo-production
fixed  target experiments.  Currently  active experiments  at $e^+e^-$
colliders are KLOE-2 at DA$\phi$NE~\cite{AmelinoCamelia:2010me} and BESIII
at  BEPC~\cite{Li:2009jd}:  the  $\eta$ and  $\eta'$ mesons  then result from
radiative decays of $\phi$ or  $J/\psi$ mesons.  The $\pi^0$, \et\ and
\etp\ mesons  can  copiously be produced in  $\gamma p$, $pp$  or $pd$
interactions  not  far from  the  production  thresholds.  The  mesons
originate  from  decays  of  nucleon  isobars (for  \et\  mainly  from
$N^*(1535)$).   There  are  two  experiments which  use  the $\gamma  p\to
p\eta^{(\prime)}$  reaction  close  to  threshold:  Crystal  Ball  at  MAMI
\cite{Starostin:2005rq}     and     Crystal     Barrel     at     ELSA
\cite{Schmieden:2007ix}.
%%%%%%%%%%%%%%%%%%%%%%%%%%
\begin{table}[htb]
\caption{\label{tagging} Production of \et\  and \etp\ mesons close to
threshold;  $p_{\mbox{thr}}$  is the beam momentum  at threshold;  $Q$
is the CMS excess energy corresponding to the maximum or optimal cross section
($\sigma$).  The  last column indicates the total  inclusive cross section
for  a given initial  state  ($\sigma_T$). The data  were extracted  from
references
\cite{Prakhov:2005qb,Mayer:1995nu,Calen:1996mn,Khoukaz:2004si,Krusche:1995nv,Plotzke:1998ua,PhysRevD.8.2789}.}
\begin{center}
\begin{tabular*}{0.9\textwidth}{@{\extracolsep{\fill}}l|rrrr}
\hline\hline
Reaction&$p_{\mbox{thr}}$&$Q$\ \ &$\sigma$ \ \ \ \ &$\sigma_T$ \ \ \\
& [GeV/c]        &{\small[MeV]}&                     &[mb]\\
\hline
$pp\to pp\pi^0$     & 0.777&122&1.3 mb&40    \\
$pp\to pp\eta$      & 1.981&40&5 $\mu$b&40    \\
$pp\to pp\eta'$     & 3.208&45&300 nb&40\\
$pd\to ^3$He$\eta$  & 1.569&2&400 nb&80\\
$pd\to ^3$He$\eta'$ & 2.434&60&1 nb&80\\
%\hline
$\pi^- p\to n\eta$  & 0.684&36&2.6 mb&50\\
$\pi^- p\to n\eta'$ & 1.432&100&100 $\mu$b&35\\
$\gamma p\to p\eta$ & 0.706&60&16 $\mu$b&0.30\\
$\gamma p\to p\eta'$& 1.447&40&1 $\mu$b&0.15\\
\hline\hline
\end{tabular*}
\end{center}
\vspace*{-0.2cm}
\end{table}
%%%%%%%%%%%%%%%%%%%%%%%%%%%%%%%%%
The  $pp\to pp\eta$  reactions  provides the  highest  useful rate  of
tagged \et\  mesons enabling studies of  the rare decays  which have a
distinct signature as, {\it e.g.}, the decay $\et\to\elp\elm$ where an
integrated luminosity  corresponding to \EA{10} \et s  is needed.  The
\et\  and \etp\  production in  $pp$  and $pd$  interactions have  the
lowest  ratio  of  the  cross  section to  the  total  inclusive  one.
However,  the signal-to-background  ratio can  be enhanced  by working
close to threshold  at the price of a  lower production cross section.
The  ideal reaction in  this respect  is $pd\to^3$He$\eta$,  where the
cross section rises very quickly, reaching a plateau already at an excess
energy  of  about  1\,MeV  above  threshold~\cite{Mayer:1995nu}.   The
presence of a doubly charged, heavy $^3$He ion in the exit channel can
easily be detected and thus  provides a very efficient and high-purity
trigger condition.  Close to threshold  the ions can be filtered using
a magnetic spectrometer with the acceptance below one degree.

A feature of the close-to-threshold experiments is that particles from
the production  processes are emitted  in a small forward  cone. Their
momenta can be  identified and measured in a  dedicated detector which
covers a limited range of  scattering angles.  The meson production is
identified (tagged)  by the missing  mass.  When the beam  momentum is
known precisely the missing mass is kinematically constrained close to
threshold and  the resolution depends  weakly on the precision  of the
kinetic  energy  determination  of  the  particles.   For  example  at
light-ion storage  rings, the beam-momentum resolution is  of the order
of 0.1\%  FWHM and a $pp$  missing-mass resolution of  typically a few
MeV/c$^2$  FWHM is achieved.   The light  decay particles  are emitted
more isotropically (since the velocity of the center of mass system is
not very high) and their  registration requires a detector with nearly
4$\pi$ sr coverage.  A typical  resolution for the invariant masses of
the decay  products is a few  ten MeV/c$^2$.  The  clear separation of
the phase space  regions for tagging and decay  particles in 
the close-to-threshold  tagging   helps,  {\it  e.g.},  in  the
determination      of      the      absolute     branching      ratios
$\Gamma_i/\Gamma_\mathrm{tot}$.

%Comparison to $\gamma p $ experiments at MAMI average luminosity \EA{30}\cms.

%Data for $\gamma p\to n\eta$ \cite{Plotzke:1998ua}

There  are  three detectors  which  have  started  a second  round  of
$\eta^{(\prime)}$ decay experiments: Crystal Ball, Crystal Barrel and WASA.
The Crystal Ball and Crystal Barrel detectors are now used for $\gamma
p\to \eta^{(\prime)} n$  experiments.  The third detector, WASA,  was build at
TSL Uppsala  where $\eta$ decay  experiments were carried out  in $pp$
and  $pd$ interactions  until 2005  at the  CELSIUS light  ion storage
ring.

The  design  of  the Crystal  Ball  \cite{Bloom:1983pc},  Crystal  Barrel
\cite{Aker:1992ny}  and   WASA  \cite{Bargholtz:2008ze}  detectors  is
similar.   The main  part consists  of a  multi-segmented calorimeter
of  NaI, CsI(Tl) and CsI(Na), respectively.  The detectors
are compact enough  so that they can be transported  to other accelerators.
In the  present configuration the Crystal Ball and  Crystal Barrel set-ups
are extended  by forward  calorimeters consisting of  BaF$_2$ crystals
from the TAPS  setup \cite{Novotny:1991aa}.  On  the other hand  they have
limited capabilities for a measurement  of charged decay products.

WASA is  the most complicated of  the three detectors.   It includes a
novel pellet target system allowing  for low background and wide angle
detection.    The   design  of   the   detector   was  optimized   for
$\pi^0,\eta\to  e^+e^-$ decays.   In addition  to  the electromagnetic
calorimeter,   the   central  part   of   the   detector  includes   a
superconducting solenoid  and a  cylindrical mini drift  chamber (MDC)
with 17 layers of  thin-walled (25$\mu$m) aluminized mylar tubes.  The
complete  detector  system was  transported  to the  Forschungszentrum
J\"ulich  in  2005,  installed  at  COSY  storage  ring,  and  it  was
operational already after one year \cite{Adam:2004ch}.  The relocation
of the detector had strengthened the experimental programme since COSY
allows for higher energy beams  (up to above $pp\to pp\phi$ threshold)
and  for  polarization.   The  detector  system was  upgraded  with  a
completely new readout  system, the refurbishing of old scintillator elements,
and an  extension of the forward  part of the detector  for the higher
energies.   The  maximum   useful  luminosity   of  the   facility  is
about $10^{32}$cm$^{-2}$s$^{-1}$.

KLOE-2, WASA-at-COSY and Crystal Ball aim at a significant improvement
of the sensitivity  of the discrete symmetries tests  in the decays of
the $\eta$  and $\eta'$ mesons  beyond the presently  achieved limits.
With an expected number of about $10^9-10^{10}$ $\eta$ mesons tagged,
a  significant   improvement  is   expected.    In  the nearest   future,
WASA-at-COSY  will provide  an order  of magnitude  improvement  of the 
branching ratio
limit for the $\eta\to  e^+e^-$ decay and the first  investigations of the
$\eta\to\pi^0e^+e^-$  decay in  the  low electron--positron  invariant
mass region, $M(e^+e^-)<120$ MeV.   High statistics measurements of the
$\pi^0$  meson   decays  are  also  planned.    The KLOE-2  experiment  at
DA$\phi$NE presented at this conference\footnote{S.  Miscetti in DISCRETE2010
  proceedings.}  is  starting data taking  and the goal is  to collect
about $10^9$  $\eta$ and $10^7$ $\eta'$ events.   With regard to the perspectives
of other  experiments, one should not forget to mention the inclusive dimuon  
spectrum at 7 TeV from the CMS experiment
(at 40  pb$^{-1}$), presented
by G.  Rolandi  at this conference.
The  spectrum shows a clear  $\eta\to\mu^+\mu^-$ peak
and  thus the  feasibility  of a  precise  BR measurement  using  the  same
technique  as previously the  NA60 experiment  \cite{Arnaldi:2009wb}.  The
first results  on $\eta'$ decays from the BESIII  experiment were recently
published \cite{Ablikim:2010kp}.

%%%%%%%%%%%%%%%%%%%%%%%%%%%%%%%%%%%%%%%%%%%%%%%%
\ack

The authors gratefully acknowledge discussions with Christoph Hanhart, Bastian Kubis and Feng-Kun Guo.
This work was  in part supported by the  European Commission under the
7th Framework Programme  through the `Research Infrastructures' action
of  the  `Capacities'   Programme:  FP7-INFRASTRUCTURES-2008-1,  Grant
Agreement N.\,227431.

%%%%%%%%%%%%%%%%%%%%%%%%%%%%%%%%%%%%%%%%%%%%%%%%

\section*{References}
\bibliography{etasym_v2}
\bibliographystyle{iopart-num_v2}

\end{document}